# 100 Years of Deuterostomia (Grobben, 1908): Cladogenetic and Anagenetic Relations within the Notoneuralia Domain

Michael Gudo[1] & Tareq Syed[2]

## Abstract

Results from molecular systematics and comparative developmental genetics changed the picture of metazoan and especially bilaterian radiation. According to this "new animal phylogeny" (introduced by Adoutte et al. 1999/2000), Grobben´s (1908) widely favoured protostome-deuterostome division of the Bilateria can be upheld, but only with major rearrangements within these superphyla. On the cladogenetic level, the Protostomia are split into two unexpected subgroups, the Lophotrochozoa and Ecdysozoa. The deuterostomes are split into the subgroups Chordata and Ambulacraria, which is not novel since Grobben (1908) introduced the Deuterostomia in this way (together with the Chaetognatha as a third line).

However, many details of the new deuterostome phylogeny do not fit traditional, morphology-based reconstructions. As a consequence, three relatively unexpected proposals for early deuterostome evolution are favoured today: An ambulacraria-scenario, a xenoturbellid-scenario, and a chordate-scenario. The first two proposals are often discussed in the literature, while the chordate-scenario is almost completely neglected. Therefore, the paper presented focuses on the chordate scenario, i.e. the hypothesis of an acrania-like "ur-deuterostomian". It is argued that the "acrania-hypothesis" is clearly preferable when biomechanic options of a polysegmented, hydroskeletal body plan are taken into account. The so called hydroskeleton hypothesis, rooted in the work of W. F. Gutmann, is the most detailed anagenetic scenario which depicts an acrania-like ur-deuterostome. Moreover, it is the only morphology-based hypothesis which is in line with all of the unexpected molecular results of deuterostome evolution (i.e. pterobranchs, echinoderms and tunicates as highly derived forms which secondarily lost the ancestral polysegmentation).

## Zusammenfassung

Seit der wiederholten Bestätigung der "new animal phylogeny" (einer molekularsystematischen, von Adoutte et al. 1999/2000 ermittelten Großphylogenie des Tierreiches) ist klar geworden, dass innerhalb der traditionellen Großeinteilung der Bilateria in Proto- und Deuterostomia (Grobben 1908) einige Änderungen vorgenommen werden müssen. Kennzeichnend für die Protostomia ist eine Neueinteilung in Ecdysozoa und Lophotrochozoa, kennzeichnend für die Deuterostomia eine Einteilung in Chordata und Ambulacraria – letztere wurde allerdings in sehr ähnlicher Form ebenfalls von Grobben (1908) postuliert.

Viele Einzelresultate innerhalb dieser molekularsystematischen Großeinteilung widersprechen morphologischen Rekonstruktionen. Für die Rekonstruktion der Deuterostomier-Evolution hat dies zur Folge, dass nur noch von drei möglichen Szenarien ausgegangen wird, welche früher allesamt als Außenseiterpositionen galten. Es sind dies ein "Ambulacraria-Szenario", ein "Xenoturbelliden-Szenario" und ein "Chordaten-Szenario". Das Chordaten-Szenario ist dabei in der neueren Literatur kaum detailliert worden. Im vorliegenden Artikel wird dies mit Rückgriff auf die sogenannte Hydroskelett-Theorie getan, die auf Arbeiten von W. F. Gutmann basiert. Das dort ermittelte Szenario eines Acranier-artigen "Ur-Deuterostomiers" hat zwei beachtenswerte Vorteile: Erstens nimmt es sämtliche unerwarteten molekularbiologischen Einzelresultate für die Deuterostomier vorweg (Hemichordaten, Echinodermen und Tunicaten als hochabgeleitete Formen), zweitens ist es im Laufe von über vierzig Jahren detaillierter ausgearbeitet worden als andere Modelle zur Deuterostomier-Evolution. Aus diesen Gründen ist es auch für eine Re-Interpretation fossiler Formen von Interesse (im Text exemplarisch für die Vetulicolia und frühe kambrische Chordaten demonstriert).

---

1) Dr. Michael Gudo, Morphisto Evolutionsforschung und Anwendung GmbH, Senckenberganlage 25, 60325 Frankfurt am Main, Germany, e-mail: mgudo@morphisto.de

2) Dr. Tareq Syed, Morphisto Evolutionsforschung und Anwendung GmbH, Senckenberganlage 25, 60325 Frankfurt am Main, Germany, e-mail: tsyed@morphisto.de

# 1. Introduction

Since KARL GROBBEN introduced the superphylum Deuterostomia one hundred years ago (GROBBEN 1908), there have been intense debates about validity, members and the exact phylogenesis of this group. Today, molecular systematics and developmental genetics offer previously unavailable clues which, however, seem to conflict with almost all morphology-based interpretations of the Deuterostomia (for a review see GEE 2001, and also WINCHELL et al. 2002). The vast majority of these morphology-based reconstructions rely on comparisons of isolated characters instead of functional character complexes. To resolve some of the ambiguities, the present paper aims at two goals. First, and most important, we will demonstrate that phylogenetic reconstructions accomplished by a function-focused approach called constructional morphology (German "Konstruktionsmorphologie"; SCHMIDT-KITTLER & VOGEL 1991; GUTMANN 1993; GUDO et al. 2002) are in much better agreement with molecular data than results from structure-based, morphological analyses. Second, we will emphasize that methods which reconstruct anagenetic pathways (i.e. successive morphological transformation steps during evolution; RENSCH 1972) should be based on independent structural-functional approaches and not on mere character sorting – however sophisticated methodologically – as in cladistic reconstructions (pure genealogies). Such anagenetic step-by-step reconstructions are potentially helpful to interprete fossil problematica (which will casually be demonstrated here with regard to the cambrian Vetulicolia (Deuterostomia incerta sedis) and *Haikouella*-fossils).

# 2. The "New Animal Phylogeny": Results and conflicts with traditional phylogenies

Over the past ten years, molecular phylogenetics changed textbook views on metazoan bauplan evolution. In the "New Animal Phylogeny" (NAP) introduced by ADOUTTE et al. (1999, 2000), traditional major domains of the animal kingdom were rearranged, apparently well-established groups split, and commonly accepted evolutionary assumptions rejected (fig. 1). On the cladogenetic level, the bilateria were devided into three superphyla, the Deuterostomia, Ecdysozoa, and Lophotrochozoa; later, the Acoelomorpha were added as an early side branch (COOK et al. 2004; JIMÉNEZ-GURI 2006). On the anagenetic level, the possibility of a polymer-segmented urbilaterian (probably growing by "terminal addition" during embryogenesis; JACOBS et al., 2005) gained increasing interest, mainly on the basis of comparative studies of the Hox gene system (e.g. BALAVOINE et al. 2002; PRUD'HOMME et al. 2003; SEO et al. 2004; DEROSA et al. 2005). This would imply that the last common ancestor of deuterostomes, ecdysozoans and lophotrochozoans can be envisaged as a coelomate annelid-like animal, and that in many bilaterian lines coelom as well as segments became lost secondarily (BALAVOINE & ADOUTTE 2003). For at least four reasons, this new view is in conflict with almost all morphology-based phylogenies: (1) Platyhelminthes split in a quite unexpected way (Acoelomorpha very basal, all other Platyhelminthes very derived; fig. 1); (2) Lophophorata neither are closely related to deuterostomes nor do they represent basal bilaterians (they are not even monophyletic, HALANYCH 2004, DUNN et al. 2008); (3) Annelida and Arthropoda can no longer be united in a taxon Articulata; (4) the bilaterian tree does not imply an overall increase of "morphological complexity", i.e. gradual transformation lines such as acoelomate=>pseudocoelomate=>coelomate (anagenesis of a secondary body cavity) or amer=>oligomer=>metamer/polymer (anagenesis of a polysegmented trunk; traditionally rated as parallel events in Chordata and Articulata) are not supported by the NAP (ADOUTTE et al. 1999).

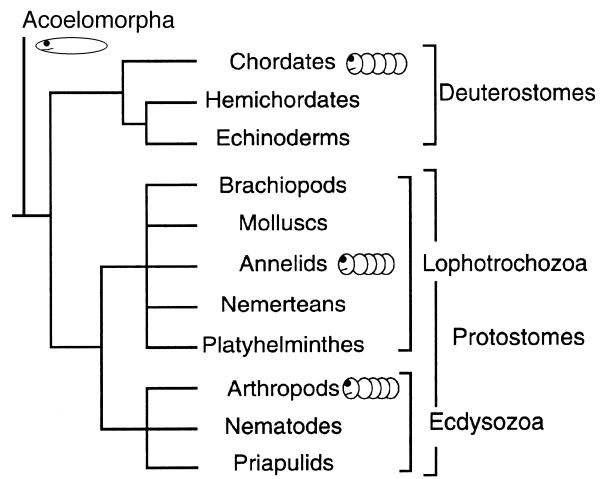

Fig. 1. A condensed view on the main cladogenetic and anagenetic aspects of the new animal phylogeny (modified from PRUD'HOMME et al. 2003; Acoelomorpha added according to COOK et al. 2004, and JIMÉNEZ-GURI et al. 2006). On the cladogenetic level, molecular systematics confirmed the traditional group of the Deuterostomia, however, Chaetognatha (arrow worms, not shown) and Lophophorata (comp. position of brachiopods) are now excluded from this taxon. Lophophorata group within the new protostome-superphylum Lophotrochozoa. The remaining protostomes are phyla of periodically moulting animals, the Ecdysozoa. On the anagenetic level, similarities in the genetic control of body segmentation suggest a polysegmented urbilaterian and loss of this feature in all bilaterian groups except the Acoelomorpha, which represent an early and isolated side branch.

## 2.1 Consequences of the new protostome phylogeny

The origin of the NAP can be traced back to the introduction of the new protostome superphyla Lophotrochozoa and Ecdysozoa in 1995 and 1997, respectively (comp. reviews in ADOUTTE et al. 1999, 2000; and HALANYCH 2004). After the exclusion from the Deuterostomia of the Chaetognatha, Pogonophora and the three lophophorate phyla (Phoronida, Brachiopoda, and Bryozoa), there now are far more protostome than deuterostome lines. GERHART (2006, p.678) recognizes 25 protostome phyla, whereas deuterostomes consist of only 6 phyla (fig. 2b and next section). Here we refrain from exploring protostome radiation; molecular analyses of protostome relationships will be mentioned only if they bear on our subsequent discussion of deuterostome phylogeny.

Obviously, if the anagenetic scenario depicted in fig. 1 is correct, then an annelid-like body plan (gradually achieved from a common, unsegmented ancestor of acoelomorphs and early annelids, see also fig. 3) will be the key to early deuterostome evolution. Some supporters of the new phylogeny, though, doubt the significance of the annelid body plan (e.g. HALANYCH 2004, p.245). However, they neglect the striking similarities in the structure and function of Hox-genes and other "segmentation genes" that function as developmental regulators on a higher hierarchical level (see for example PRUD'HOMME et al. 2003 and DEROSA et al. 2005) without offering an evolutionary explanation. Another reason for rejecting the hypothesis of a polysegmented urbilaterian is that this scenario implies a multiple, "non-parsimonious" loss of segmentation in several bilaterian lines (e.g. JENNER, 1999; comp. also fig. 1). On the other hand, BALAVOINE & ADOUTTE (2003) pointed out that many of these non-segmented phyla possess characters that can be interpreted as traces of a former segmentation, and that comparative developmental genetics should be applied for further clarification (as in the case of the widespread "terminal addition" mode of growing, JACOBS et al. 2005).

A significant feature of the "polysegmented urprotostomian" concept is that only in this case the classical anagenetic model of arthropods evolving from annelid-like precursors can be upheld, despite the abolition of the taxon Articulata. The "polysegmented urprotostomian" concept also implies the derivation of the Acrania and Craniota from annelid-like precursors, an old idea which has often been critized by pointing at the ventrally localized nervous system of annelids (gastroneural condition), which contrasts with the dorsal nerve tube of the Acrania and Craniota (notoneural condition). However, this argument is valid only if two currently existing body plans are forced into a phylogenetic sequence (for example, by turning annelids upside down to reach the chordate-condition), a procedure which is anti-evolutionistic and forms the root of countless misconceptions in evolutionary biology. In our reconstruction, we prefer to follow RUPPERT (2005, p.12) and explore the implications and consequences of the assumption that a common ancestor of Gastroneuralia and Notoneuralia was a polysegmented animal *without* a highly specialized nervous system. Also LOWE (2008: 1575) points out "that a hypothetical ancestor was not necessarily characterized by a central nervous system" (which of course does not exclude the possibility of a common ancestry of bilaterian nervous system centralization, see DENES et al. 2007). Quite obviously, distinct architectures of the neural systems are possible at the end-points of different anagenetic pathways that originate from this common ancestor (thus, by using HATSCHEK'S (1891) terms Gastroneuralia and Notoneuralia we refer to existing groups rather than their ancestors). In this case, the question of deuterostome origins becomes linked to the question of the gradual evolution of the notoneural condition. It can best be answered by discussing probable evolutionary advantages of a dorsal muscle concentration in the line that leads to the earliest deuterostomes, which in turn requires sound biomechanical reconstructions of hypothetical ancestors (the constructional approach; see section 3).

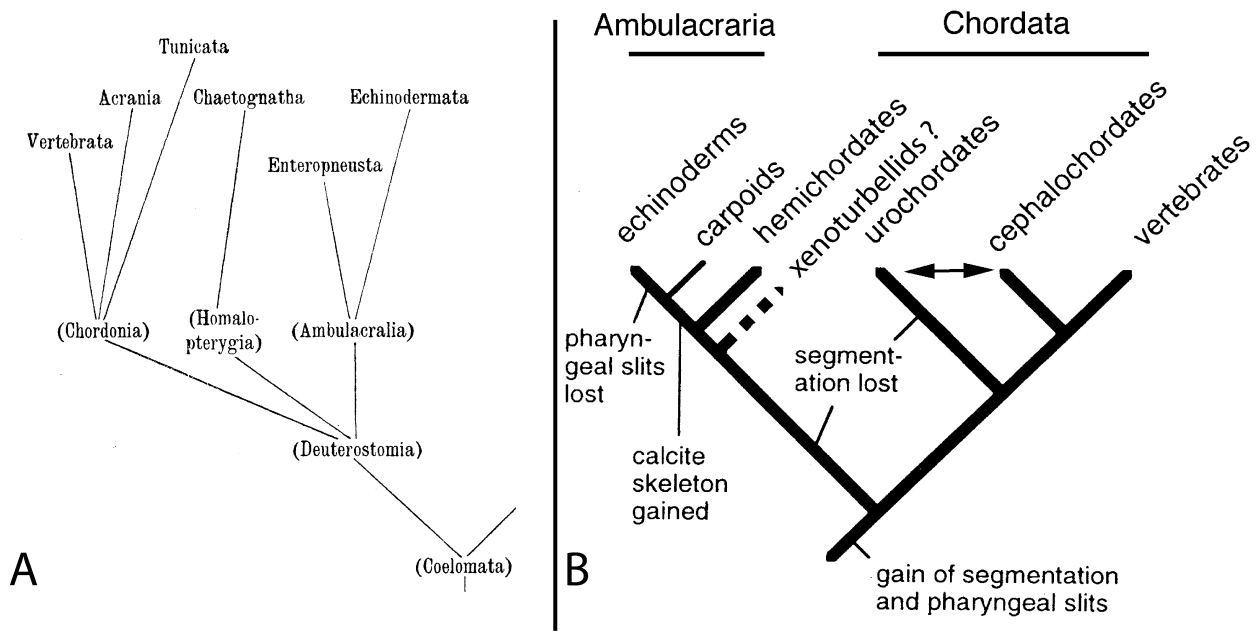

**Fig. 2.** Two distinct but partly similar views on deuterostome cladogenesis and anagenesis. 2A. Karl GROBBEN introduced the term "Deuterostomia" for the subgroups Chordonia, Ambulacralia and Chaetognatha. Deuterostomia and Protostomia (not shown) originate from a common coelomate precursor (detail of a phylogenetic tree from GROBBEN 1908). 2B. Molecular systematics reanimated the often rejected idea of a Chordata-Ambulacraria dichotomy in the Deuterostomia. In contrast to 2.A., Chaetognatha are excluded from the Deuterostomia; instead, the monotypic phylum Xenoturbellida is now regarded as a basal branch of the Ambulacraria. Following fig. 1A, a segmented, probably coelomate ur-deuterostomian is suggested. Gill slits are present in ur-deuterostomes, and are secondarily lost in the Xenoturbellida and recent echinoderms (carpoids are extinct echinoderms with gill slits). The double arrow indicates that the position of Tunicata (urochordates) and Acrania (cephalochordates) has to be changed according to newer results. Independent of these modifications, the scheme implies a secondary loss of segmentation in tunicates, and also in xenoturbellids and echinoderms (modified from GEE 2001; Xenoturbellida added following BOURLAT et al. 2006).

## 2.2 Consequences of the new deuterostome phylogeny

Molecular systematics of the Deuterostomia unanimously support the division of this superphylum into Chordata and Ambulacraria (fig. 2b; see Halanych 2004 for review). Together with the exclusion of the lophophorate phyla from the deuterostomes, this "new" picture somehow resembles Grobben´s (1908) introduction of the Deuterostomia (fig. 2a). However, Grobben´s inclusion of the Chaetognatha (arrow worms) into the Deuterostomia is now refuted by molecular results (Halanych 2004; Marlétaz et al. 2006, Dunn et al. 2008). Instead, the flatworm-like *Xenoturbella bocki* became a new member of the deuterostomes, probably as the basic ambulacrarian line; this rearrangement contradicts former views of echinoderms as basal ambulacraria (comp. fig. 3b and Stach et al. 2005; Bourlat et al. 2006; Dunn et al. 2008, but also Perseke et al. 2007). Another surprise came with the molecular phylogeny of the hemichordates, because here the Pterobranchia appear as a highly derived line (Cameron et al., 2000) that probably originated from enteropneust-like precursors; this reverses the relationship favored by most morphology-based phylogenies. Similarly, the division pattern along the Chordata-branch was rearranged (comp. fig. 2b). Instead of the previously almost universally accepted Acrania/Craniota taxon, there now is support for a sister-group relationship between Tunicata and Craniota (Philippe et al. 2005; Bourlat et al. 2006; Delsuc et al. 2006; Wada et al. 2006, Dunn et al. 2008; for a contradicting interpretation, see Mallatt & Winchell 2007). In any case, tunicates now are seen as a highly specialized side branch, and no longer as representatives of the ancestral chordates, due to their reduced genome size and dispersed Hox cluster (e.g. Ikuta et al. 2004; Seo et al. 2004; Hughes & Friedman 2005). The derived status of the Pterobranchia and Tunicata leaves Xenoturbellida, Enteropneusta, and Acrania as the only possible representatives of the earliest deuterostome line. In the context of the idea of a polysegmented ancestor, the Acrania would appear the best candidates for relatively basal deuterostomes due to their polysegmented body (chorda-myomere system). However, many developmental biologists favour an enteropneust-like condition at the basis of the clade, arguing that the dorsoventral expression pattern of bmp and chordin genes is protostome-like in enteropneusts (e.g. Gerhart 2006, Lowe 2008). If this assumption is correct, the polysegmentation of the urbilaterian became reduced during evolution of an enteropneust-like ur-deuterostomian, followed by a "re-polymerization"-process in the line leading to the chordates. In the alternative scenario, an acrania-like ur-deuterostomian evolved from the polysegmented urbilaterian, and polysegmentation vanished in the lines leading to the ambulacraria and tunicata (fig. 2b). This would imply that the protostome-like bmp-chordin-axis of the enteropneusts is a case of evolutionary parallelism that probably occurred during a process of strong simplification and modification. Both of these scenarios have their merits, at least from a developmental biologist´s point of view, and further studies must be awaited (see also section 5.3). For example, if a bmp-chordin-axis could be demonstrated in the xenoturbellids, with the mouth located at the chordin side (as in craniotes), this could weaken the view that the enteropneust condition is ancestral in deuterostomes (obviously, this argument also depends on the systematic position of *Xenoturbella*, which is relatively uncertain at the moment – comp. 5.1. for details).

There appears to be agreement regarding the ancestral deuterostome organization in one crucial point: nearly all developmental genes so far studied in comparative analysis of enteropneusts and acrania/craniota are not expressed in larval or juvenile stages; their activation starts when reaching the adult condition. Thus, as Gerhart (2006, p. 681) states, the ancestral larva is a poor candidate for ur-deuterostomian organization, compared with the ancestral adult. In consequence, the various "neotene larva"-scenarios for deuterostome evolution (e.g. Salvini-Plawen 1998) are rejected (see also Lacalli 2005), and anagenetic reconstructions have to focus on stepwise biomechanical modifications of adult forms.

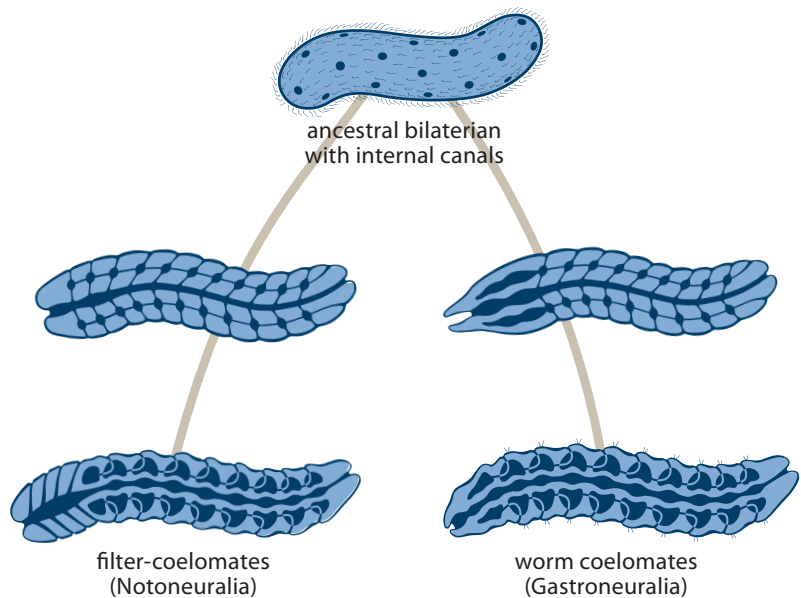

**Fig. 3.** Functional design of the ur-deuterostomes and ur-protostomes. Here, the last common ancestor of deuterostomes and protostomes is reconstructed as a worm-like animal with a metameric hydraulic skeleton functioning as force absorber and force transmitter (evolving from a compact, gelatinous bilaterian with inner canal systems). Left side: Through specific coordinations of the activity of circular and longitudinal muscles, lateral undulations could be generated which resulted in swimming movements. The cranial region of this ancestral animal was designed as a filtering device, giving rise to the branchial apparatus which is a common structure in deuterostomes ("filter coelomates"). Right side: The alternative design is presented by creeping, benthic forms which developed a sucking pharynx for taking up particles from the substratum ("worm coelomates"). The pelagic forms developed a dorsally concentrated nervous system (innervation of muscles around the chorda dorsalis, comp. fig. 4), while in the benthic forms a ventral concentration of nerves took place (probably in connection with parapodial movement).

## 3. Anagenetic reconstructions by Konstruktionsmorphologie

Anagenetic reconstructions should present evolutionary scenarios that show how the functional design of organismic body structures can be transformed without gaps in functionality. Plausible scenarios can be established by ruling out dysfunctional intermediates, and by explaining how the proposed pathways of transformation are canalized and constrained. In the quasi-engineering approach of "Konstruktionsmorphologie", the structures of living organisms are treated as biotechnical units. This conceptualization of plausible reconstructions is not comparable with anagenetic hypotheses established by the mapping of characters on genealogies; it does not even require a cladogenetic working hypothesis. Instead, a detailed analysis of the functional design of recent organisms is the starting point for a "retro-engineering" evolutionary interpretation. Here, an initial analytical dissection phase identifies and describes the structural elements and analyses their mechanical roles and mutual interdependence. A subsequent integrative phase uses these data to create a mechanically coherent model of the organismic apparatus. It takes into account the biomechanical and material properties of the various tissues and fluid fillings which constitute every organism. Thus, organisms are interpreted as "hydraulic machines" in this approach. Konstruktionsmorphologie originated in a series of papers which aimed to reconstruct vertebrate anagenesis from invertebrate precursors (GUTMANN 1967a et mult.). The evolutionary modifications of hydroskeletal body plans are a central aspect in invertebrate-vertebrate transformation models, and these anagenetic models automatically affect the understanding of deuterostome radiation.

## 4. Evolutionary Scenario for Deuterostomes

This section presents the basic reconstructions for the Notoneuralia domain, as elaborated by the approach of Konstruktionsmorphologie over the past forty years. The hypothesis of a polysegmented hydroskeletal Urdeuterostomian, which gives rise to a chordate branch on the one hand and an ambulacrarian branch on the other, has been revised and detailed many times (e.g. GUDO & GRASSHOFF 2002). Central arguments of the latest version are summarized in the following, offering a conclusive morphological interpretation of deuterostome molecular phylogeny. Moreover, we think that this is one of the most detailed scenarios of deuterostome evolution. As such it is of special interest for palaeontologists, because fossil findings can be interpreted with regard to the hypothetical intermediates described in the step-by-step-reconstructions.

### 4.1 The Ur-Deuterostomian

The first aspect we have to deal with is the body structure of the supposed deuterostome ancestor, i.e. the Ur-Deuterostome and the Ur-Bilaterian. The polysegmentation of these forms was proposed by GUTMANN (1966, 1967a, 1972a,b), now gaining support from developmental genetics (fig. 1). By the retro-engineering approach, the body structure of the deuterostome ancestor can be outlined in more detail, i.e., the interaction of anatomical and histological structures is described: It was an elongated worm-like animal with serially arranged coelomic cavities that served as a hydraulic skeleton (comp. fig. 3). Lateral undulations, one of several types of body movements possible in such a construction, were performed by muscular contraction and antagonistic interaction transmitted by the fluid filling of the coelom. The coelom is one of several differentiations of former fluid-filled canals lined with a ciliated epithelium. One of these canals functioned as intestinal tract, others functioned in the distribution of nutrients and the excretion of metabolic products. Additional canals might have been present, but speculations about their number and function lies beyond the possibilities of our reconstruction. One specific longitudinal canal, located dorsally of the intestinal tract, will be of special importance in the subsequent argumentation (shown in fig 4).

Biomechanicly, the lateral, serial arrangement of hydraulic cavities minimizes internal friction during body movements. When the body bends laterally, the fluid in these proposed coelomic cavities is deformed by the muscles of one body side and thereby stretches the muscles of the opposite side. The tissues that separate the chambers of this segmented hydraulic skeleton constrain the deformation of the fluid and permit smooth movement. We will assume a state in which, similar to living annelids, there were two types of internal walls that bordered coelom chambers. First, there were the transverse septa oriented perpendicular to the long body axis, which are called dissepiments. Second, septa along a plane parallel with the long body axis divided the serial body cavities into two lateral portions, thus establishing a truely bilateral biomechanic architecture without which it would be problematic to define lateral undulation as a specific mode of motility (HERKNER 1991). Such a longitudinally oriented septum is called mesenterium. It should be noted that while this hypothetical annelid-like urbilaterian resembles modern annelids biomechanicly, there is no requirement for body tissues to be as highly differentiated as in recent annelids. Assuming that our urbilaterian originated from gelatinous early metazoa (fig. 3), the body can be assumed to have consisted of relatively undifferentiated gelatinous tissues with contractile cells and a grid of connective fibres (the latter being components of the extracellular matrix). The potential of such "gelatinous fibrous tissues" or, in German, "Gallertfaser-Gewebe", for evolutionary differentiation has been discussed elsewhere (GRASSHOFF & GUDO 2001, 2002).

The fluid-filled body works as a hydrostatic skeleton (CLARK 1964), which means that forces exerted by contracting muscles are transmitted to antagonistic muscle groups by the volume-invariant deformation of fluid-filled cavities. In such a hydraulic softbody system performing lateral undulations, the hydraulic system transmits the muscular forces between the left and right body side, which contract alternatingly. Under contraction of the longitudinal muscles alone, the body would be shortened stepwise and increase in width, but circular muscles as well as contractile elements that might be present in the dissepiments and mesenteria work against these forces (HERKNER 1991) and so the body bends, and ultimately moves forward. For physiological reasons it is necessary that each of the hydraulic units (=segments) has its own excretory system such as the metanephria in existing annelids. These metanephridia have their internal opening in one coelomic cavity and their terminal end in the adjacent (caudal) segment (fig. 4).

This organism moved by lateral undulations; a pelagial life made the active acquisition of food possible. Any morphological transformation that improved the efficiency of locomotion or reduced the amount of muscular activity required for the dynamic control of body shape during locomotion was advantegous at this comparatively unspecialized stage. In an engineering sense, the most efficient modification of this body structure would be a reduction of the mass of the circular muscles, which only function to stabilize body length during undulating movements and enforce a more or less constant body diameter. Controling body shape by muscular action alone is energy-consuming and metabolicly costly, and it limits the overall body size (GUTMANN 1972a; GUDO & GRASSHOFF 2002).

### 4.2 The Chordates

The evolutionary transformations leading to a reduction of the circular muscles have to be interpreted in the context of interactions between existing structures and of causal histogenetics. If the body length is stabilized by circular muscle activity, the tissues in the body can differentiate in specific ways. During swimming movements – performed by the lateral bending of the body via longitudinal and circular muscles – tissues are set under compressive and tensile stresses. Especially the tissues in the sagittal plane of the body have to absorb lateral and longitudinal forces (HERKNER 1991). The tissues react by histological reorganization: the connective tissue fibres and contractile cells orientate perpendicularly to the working forces. In interaction with the already existing fluid-filled canal dorsal of the intestinal tract, force-absorbing structures developed in the sagittal plane of the body.

The dorsal canal above the intestinal tract became the key structure for the evolving notoneuralian organization (GUDO & GRASSHOFF 2002). It provided a place for nervous fibres stimulating the longitudinal muscles on the left and right body flank. As a consequence, control and stimulation of locomotive muscles is not disturbed by the generated mechanical forces. In a fluid-filled canal, bending forces are low and compressive forces are compensated by the internal liquor and the enveloping connective tissue. This effect is further enhanced when the canal is closed at both ends. It represents a hydraulic system in the midline of the body, stabilizing body length at least to a certain extent. As a consequence, radial muscles lose some of their importance as control elements of body width, and can be gradually reduced. Moreover, the tissues between the neural canal and the intestinal tract could develop additional histological differentiations that enhanced their functionality as a length-stabilising yet flexible structure. As mentioned before, these tissues formed a force-absorbing structure, possessing fibres orientated perpendicular to the working forces. Finally, the notochord resulted: a connective tissue enclosing a contractile system that provided length stability (HERKNER 1991). The notochord can perform the function of length stabilization much better than the ancestral interaction of lateral muscles and the fluid-filled neural canal. Obviously, the functional design which evolved by these transformations is the body structure of an early chordate (fig. 4, for more details see GUTMANN 1972a; GUTMANN 1997; GUDO & GRASSHOFF 2002).

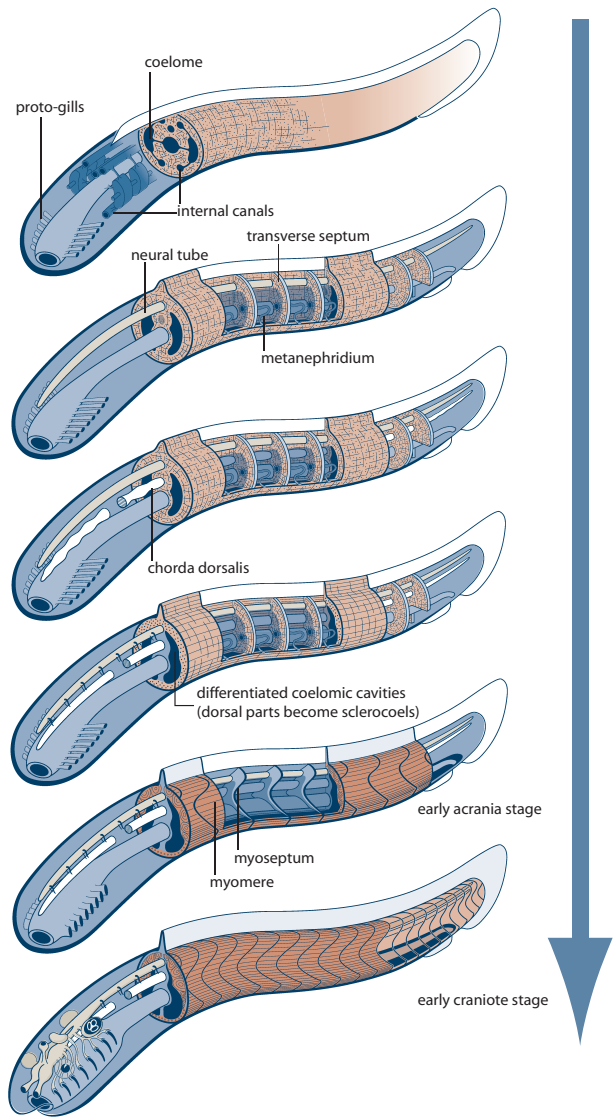

**Fig. 4.** Evolutionary scenario for the origin of chordates from polymersegmented ancestors possessing a hydraulic skeleton. The sequence starts with the hypothetical proto-deuterostome, an actively swimming, filtrating organism. The evolution of the neural tube and the chorda dorsalis are optimizations of the lateral undulating mode of swimming. Thus, deuterostome radiation reflects different options of the notoneuralian and chordate organization. Due to the dorsal concentration of muscles, the coelom and excretory system (metanephridia) are rearranged in the lineages leading to the acrania and craniota (modified from GUDO & GRASSHOFF 2002; see text for details).

This makes sense hydromechanicly as no hydraulic fluid is lost in an uncontrolled way when muscular activity sets the fluid filling under pressure (CHAPMAN 1950; CHAPMAN 1958; QUILLIN 1998).

The front end of this polysegmented animal carries a filtration apparatus; canals run from the foremost part of the gut to the lateral body wall so that water can flow in through the mouth and out through lateral openings. Nutrition particles are collected and held by mucus, and transported to the pharynx via a ciliated ridge (i.e. the putative endostyle, GUDO & GRASSHOFF 2002).

The histological configuration of such an early chordate is represented by *Branchiostoma lanceolatum* and other extant acranians. Histological investigations show the notochord as a contractile hydraulic organ surrounded by the hydraulic sclerocoels, which allow free gliding of the notochord in the muscle packages during lateral undulation (GUTMANN 1971; BONIK & GUTMANN 1978). With respect to the body cross-section, the notochord is positioned somewhat dorsal of the midpoint. This configuration leads to particular biomechanical problems for lateral undulations. Engineering considerations show: if (1) muscle packages would be oriented in a way that the generated forces work exactly parallel to the body axis, and (2) the length stabilizing, force absorbing structure would not be in the middle of the body diameter, then the horizontal undulation of the body would include a slight dorso-ventral bending, causing inefficient locomotion. But in existing animals, muscle fibres are arranged in a specific manner, viz. in a long elongated spiral that extends from the front to the hind end and the notochord in its center. This orientation of muscle fibres is attained by a reorganisation of the dissepiments (which later will evolve into myosepta) which connect the several portions of longitudinal muscles coherently with each other into the well known zig-zag pattern of recent chordates (BONIK & GUTMANN 1978). However, an alternative solution is also possible: an enlargement of the dorsal muscle packages, which results in a midline position of the chorda over the entire body length and allows the perpendicular orientation of the myosepta. This evolutionary solution is shown by some fossil early chordates, such as *Haikouella lanceolatum* or *Pikaia*, found in early cambrian sediments (BRIGGS & KEAR 1994; CHEN et al. 1995; CHEN & LI 2000; MALLATT & CHEN 2003; SHU et al. 1999, 2003). It can be interpreted as an ancestral functional design in the evolutionary field of chordates; however, these organisms represent side branches.

The transverse septa can be reduced in the region of the coelomic cavities when length stabilisation is achieved by the notochord, so that the coelomic fluid does no longer function as part of a hydraulic skeleton for lateral movements. Consequently, the tissue mass – which always comes at material and energetic costs for production and maintenance – is reduced. The result is one large body coelom, containing the intestinal tract and the internal organs. This fluid-filled cavity still functions as a hydraulic skeleton for the dorsoventral bending driven by the epaxial and hypaxial musculature. From this chordate animal, further improvements of lateral undulation open up the pathway to the craniotes (GUTMANN 1967b; GUTMANN 1972a; GUTMANN 1972b). Another anagenetic pathway originating from the simple chordate construction defined above leads to the recent acranians which preserved features of this ancestral structure, due to some physiological constraints. As discussed by SCHMITZ et al. (2000), respiration through the branchial gut contributes only up to 10 percent to the overall respiratory activity of small acranians, while the bulk of the gas exchange takes place in the tissues of the peribranchial chamber, over the entire body surface, and at the surfaces of the coelomic cavities. This explains why acranians have to preserve their locomotory apparatus although they are hemisessile organisms dwelling in sediments. The tunicates, on the other hand, represent an alternative: they enlarged their peribranchial cavity and perform respiration across its surface alone. As a consequence, they were able to reduce their muscular tail, preserved in the famous tadpole larvae. Among adults, only in the tunicate subgroup of the copelates which have no peribranchial cavity, the tail persists; the copelates are an early tunicate side branch (comp. GUDO 2004). The enigmatic Cambrian Vetulicolia fossils (SHU et al. 2001) probably represent intermediates between swimming chordates with enlarged peribranchial cavity and ascidia-like tunicate forms (comp. the predicted prototunicate in fig. 5 and fig. 8).

As an interim summary, we conclude that our postulated ancestral deuterostome gave rise to three major evolutionary lineages: the acranians, the craniotes, and the tunicates. The ancestral deuterostome was a swimming, filter-feeding chordate construction. This mode of life led to the craniotes on one hand and, as an alternative specialization, to the peribranchialchordates which divided early into the acranians and the tunicates, on the other (fig. 5).

### 4.3 Ambulacrarians

The ambulacrarians (hemichordates plus echinoderms; METSCHNIKOFF 1881) are commonly seen as primitive organisms and therefore assumed to be basal in evolutionary history. However, not alone the NAP implicates a different interpretation, but also biomechanical considerations make it plausible that ambulacrarians are highly derived animals.

Starting again with the early chordate defined above, the following scenario can be reconstructed. In the ancestral chordate, the coelomic cavities of the metameric ancestor had been preserved as sclerocoels around the notochord, facilitating practically friction-free gliding of the notochord past the lateral muscle groups when the body undulated. This organism had a peribranchial chamber, enabling a hemisessile life-style burrowed in the sediment, as in recent acranians. Modern lancelets burrow by wriggling; they 'hammer' themselves into the sediment by notochord-contractions, supported by intense lateral undulations of the trunk. However, this type of motility is energetically costly, and is only applied for short, fast movements into sandy sediments. For a more persistent burrowing in softer sediments, peristaltic body deformation would effectively support the wriggeling movements. However, such quasi-peristaltic movements can be perfomed only by the front end of the ancestral chordate, when the longitudinal muscle fibres in the oral region attain a plywood-like crisscross orientation, and when the sclerocoels around the notochord enlarge. During these transformations, the cranial end of the chordate was transformed into an organ which allowed burrowing in fine sediment by peristaltic movements. Finally, when the peristaltic organ differentiated into a proboscis and a collar, the wriggeling movements of the hindpart of the body became redundant. As a consequence, the notochord and the massive longitudinal muscles could be reduced in caudo-cranial direction. The original peribranchial chamber behind the peristaltic organ (that is, caudal of the collar), was dorsally opened, enforcing nerve fibres to rearrange in the way known from recent enteropneusts. The gonads maintained their place in the "wings" of the peribranchial chamber. The result of this transformation was an ancestral enteropneust-like organism (GUTMANN & BONIK 1979).

This organism possessed an organisation of its coelomic cavities which became important not only for the functioning of the enteropneust construction, but also for subsequent evo-lutionary transformations. A pair of trunk coeloms extended along the mesenterium into the collar region. The burrowing enteropneust moves by peristaltic creeping, anchors its collar in the sediment and contracts the longitudinal muscles of the trunk pulling it behind. In this way, the pressure in the trunk coelom including its protrusions into the collar rises. The resulting stiffening of the collar supports its anchoring in the walls of the burrowed tube (fig. 6).

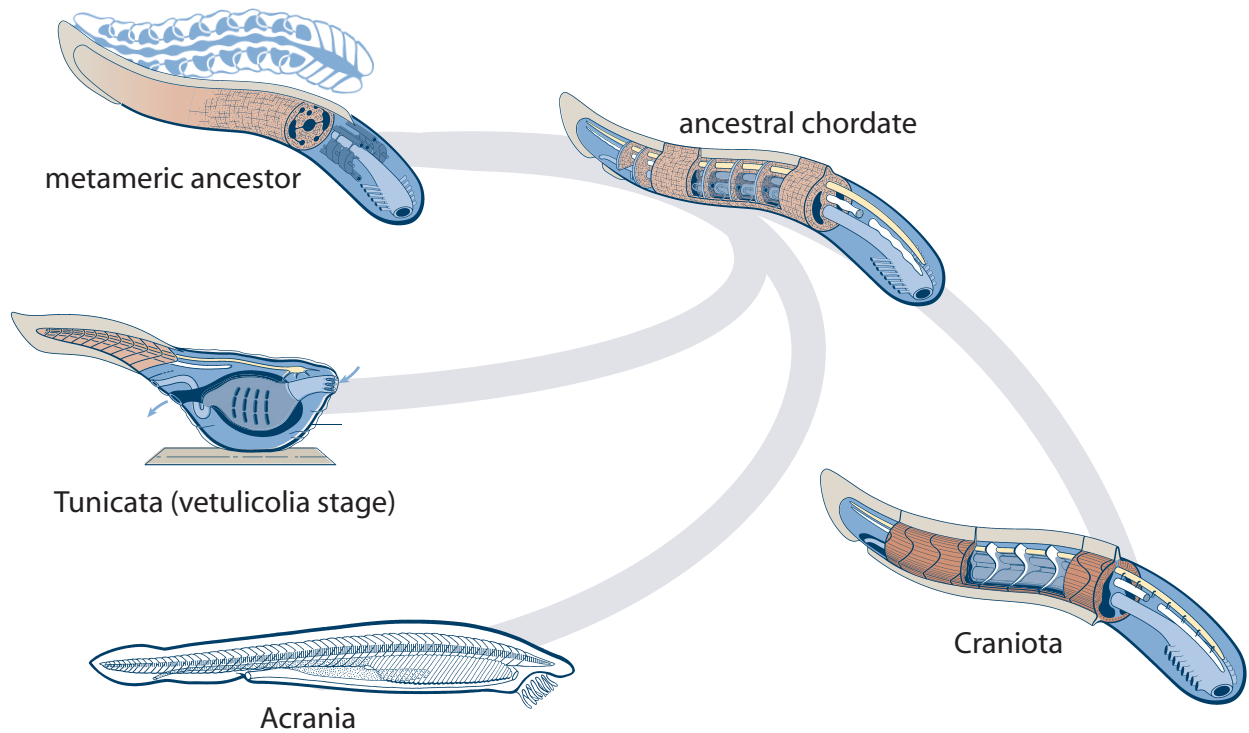

**Fig. 5.** Anagenetic relations of the chordate branch of the deuterostomes. According to biomechanical considerations, chordates appear basal, while tunicates (= urochordates) are derived from acrania-like intermediates. Note that the branching order of the Tunicata, Acrania and Craniota still remains unresolved in molecular phylogenies (comp. fig. 2b).

The reconstruction of the next evolutionary steps starting from the enteropneust organization is straight-forward (comp. fig. 7). Extensions of the collar became tentacle-like structures as found in recent pterobranchs (GUTMANN 1972a, BONIK et al. 1978, GUTMANN & BONIK 1979). The transition to echinoderms occurred in a separate lineage in which not the gut and its terminal openings were shifted in the mesentery – as it happened in the pterobranchs – but in which the gut together with the mesentery became bent into a U-shaped structure. In this way the trunk coelom was enlarged, forming a hydraulic capsule which could be pressurized continuously by muscular activity. Additional loops of gut and mesentery developed, providing internal tethering structures. These tethering structures controlled the patterned bulging of the body wall when the fluid filling was set under pressure. Due to the bending of the gut and to basic physical principles acting when hydraulic systems come into close contact, the trunk formed five bulges forcing the body into a overall pentaradial organisation. These bulges enlarged more and more and provided mechanical support for the tentacles that developed from the collar. Various complex transformations of the geometry of the coelom and the nervous system as well as histological transformations occurred which we will not discuss in detail; for more complete descriptions of the relatively straight-forward derivation of echinoderm body-plans from enteropneust-like constructions, see GUDO (2005) and GUDO & DETTMANN (2005).

With this biomechanics-based evolutionary scenario, the evolutionary history of the deuterostomes is complete. The ancestral deuterostome-construction offered the biomechanical option to develop a peristaltic organ (consisting of proboscis and collar) by simple morphological transformations. When these structures were fully developed, the typical chordate features could be reduced and the body structure of an enteropneust evolved. Thus, pterobranchs and echinoderms are highly derived hemichordates. However, it remains unclear at this stage whether pterobranchs and echinoderms are widely separated (i.e. independently from enteropneust-like predecessors) or whether they evolved from an ancestor with a body structure quite similar to pterobranchs.

## 5. Discussion

If the molecular reconstruction of deuterostome phylogeny as shown in fig. 2b is accepted, then the anagenetic scenario of the hydroskeleton hypothesis, published in a comprehensive form as early as 1966 by W. F. GUTMANN, will be of high interest as it sheds light on the causes for the contradictions between phylogenies based on structural characters as opposed to molecular ones. Provided that the cladogenetic and anagenetic implications of the NAP are correct, it appears that approaches based on morphological structures suffer from an inability to identify secondary simplification events (JENNER, 2004). In our constructional morphology approach, this is not the case.

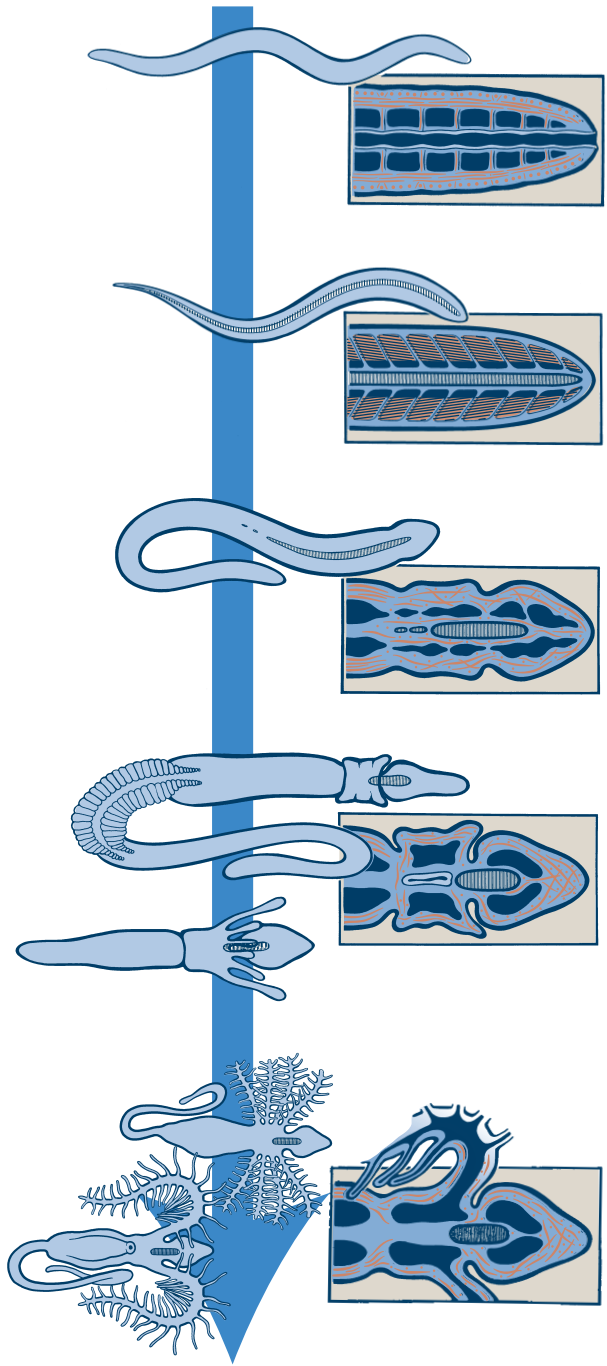

**Fig. 6.** Evolutionary transformation from chordates to ambulacrarians (modified from GUTMANN 1989; see text for details).

Decades ago, "Konstruktionsmorphologie" has interpreted tunicates, hemichordates, and echinoderms as highly derived forms featuring secondary losses of typical chordate characters, which nicely fits the molecular results shown in fig. 1 and fig. 2b. Moreover, the exclusion of the lophophorate phyla, the Chaetognatha, and the Pogonophora from the Deuterostomia is in full agreement with the hydroskeleton hypothesis (GUTMANN 1966 included the Pogonophora into the Deuterostomia, but corrected this in 1972 by describing them as derived annelids. The latter is now confirmed by molecular data, see HALANYCH 2004, p.239 f.). In particular, it is worth emphasizing that on the basis of recent molecular findings, the lophophorate phyla do not seem closely related to the Deuterostomia. The taxon "Epineuralia" consisting of Lophophorata and Deuterostomia, which formed the basis of the – now fully refuted – cladogenetic and anagenetic proposals of V. SALVINI-PLAWEN (1998) and NIELSEN (2001), appears today to be artificial. Lophophorates are highly derived protostomia, according to the hydroskeleton hypothesis; this probably is one of the most specific similarities between the hypothesis and the NAP, as noted by BALAVOINE & ADOUTTE (2003, p.142) with regard to the brachiopod derivation model of GUTMANN et al. (1978). Furthermore, the hydroskeleton hypothesis interprets the pelagic Copelata as an early branch within the Tunicata (mentioned in section 4.2), which is also supported by molecular data (MALLATT & WINCHELL 2007). Thus, it now appears that a very first "total evidence model" is possible, i.e. a complete integration of molecular results and independently elaborated morphological reconstructions (SYED 2003; SYED et al. 2007). Concerning early deuterostome radiation, it has been stated that three competing models have to be discussed: an ambulacrarian model, a chordate model, and a xenoturbellid-model (SWALLA & SMITH 2008: 1565). It is quite surprising that the chordate model is completely ignored in the newer literature, and that the scenario of the hydroskeleton-hypothesis is still the only detailed proposal here. Nevertheless, several problematic aspects of the chordate model remain to be clarified, as discussed below.

### 5.1 The unresolved position of *Xenoturbella bocki*

The possible deuterostome relationship of xenoturbellids is a feature of current molecular phylogenies which has not been subjected so far to a rigorous retro-engineering analysis in the context of the hydroskeleton hypothesis. If Xenoturbellida are an early side branch of Ambulacraria as suggested by fig. 2b, then an enteropneust-like precursor would be the most obvious starting point for an anagenetic interpretation. We do not attempt to detail such a model here, because the notion of *Xenoturbella* as a basic ambulacrarian line (BOURLAT et al. 2006) has recently been challenged by PERSEKE et al. (2007), according to whom xenoturbellids are closest relatives of all other deuterostomes, or in some analyses even a sister group of the Protostomia + Deuterostomia-clade. However, the latest results from molecular systematics, including an analysis of their Hox-repertoire, support the inclusion of Xenoturbellida into Deuterostomia, and their branching as basal Ambulacraria (FRITZSCH et al. 2007, DUNN et al. 2008). These results imply that *Xenoturbella* has to be interpreted as a secondarily simplified animal, which lost coelomic cavities and trimeric segmentation from a hemichordate-like stage. In support of this idea, it has to be stressed that according to the NAP the vast majority of flatworm-like body plans has to be interpreted as "coelomates without a coelom" (BALAVOINE, 1998), i.e. as highly derived side branches, independent of their exact branching position (comp. position of the platyhelminthes in fig. 1).

### 5.2 Echinoderms: Descendants of pterobranchs or of enteropneusts?

The hydroskeleton hypothesis has been challenged on the grounds that the NAP does not support a Pterobranchia + Echi-

17

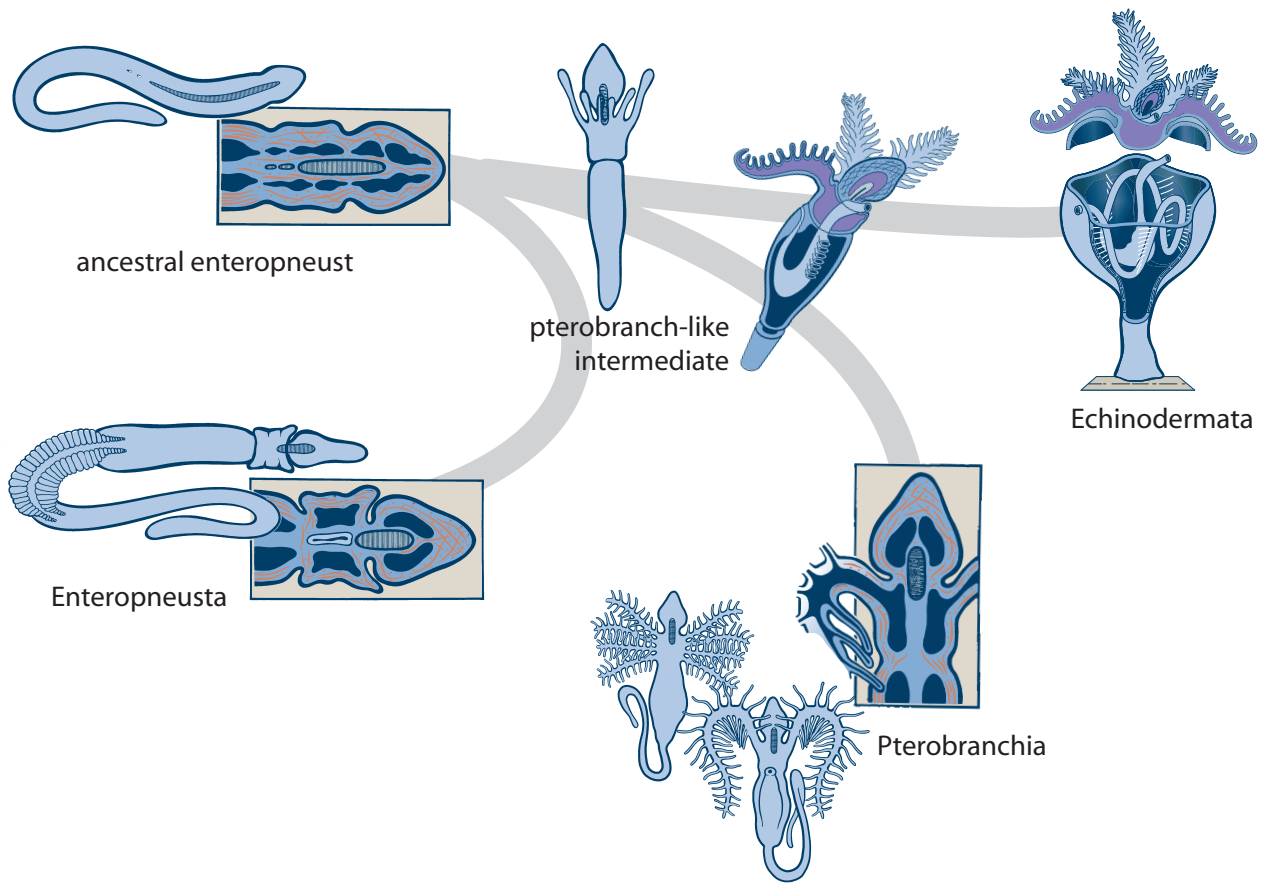

**Fig. 7.** Anagenetic relations of the ambulacrarians. According to the presented scenario (fig. 5), the functional design of enteropneusts can be derived from an ancestral chordate. From such a body structure, the recent enteropneusts, pterobranchs, and echinoderms developed as independent lineages. This does not necessarily imply that Echinodermata are the sister-group of the Pterobranchia, but rather that echinoderms have to be derived from ancestors that had a functional design similar to pterobranchs.

nodermata-clade (WINCHELL et al. 2002, p.774). This argument is not sound since the anagenetic model of echinoderm evolution as given by GUTMANN (1972a) and, in more detail, by GUDO (2005), does not necessarily imply a sister group relationship between Pterobranchia and Echinodermata. As GUTMANN (1981, p.74) explained:

> "Because the adult morphology of echinoderms is so profoundly modified from that of typical acrania-like ancestors, it is difficult to suggest a reasonable transformation series and to link possible intermediate stages with known groups. Most evidence suggests a very close link between echinoderms and hemichordates so that I will assume that the pterobranchs or at least the enteropneusts represent the chordate construction from which echinoderms evolved."

In other words, an independent evolution of pterobranchs and echinoderms from an enteropneust-like ancestor is not excluded (in fact, this possibility was explicitly demonstrated by BONIK et al. 1978). In this scenario, which fits the results of the NAP, the tentacles of pterobranchs and echinoderms evolved independently. Our figs. 7 and 8 show both possibilities: A close relationship of echinoderms and pterobranchs in fig. 7fig. 1, and a probably independent evolution of these lines in fig. 8. What both reconstructions have in common is that the tentacles of pterobranchs and echinoderms arise from extensions of the enteropneust collar (see also fig. 6). In general, the collar region of recent enteropneusts shows considerable variability as demonstrated by the bizarre, enlarged collars of the former "Lophenteropneusts", deep sea enteropneusts described by HOLLAND et al. (2005). Taken together, the reconstruction of the ambulacraria-domain does not imply one single cladogenetic tree, but rather an anagenetic scenario in which echinoderms and pterobranchs are more derived than enteropneusts. It cannot be overstressed that this long-time minority opinion is now fully supported by molecular phylogenetics.

### 5.3 The basic deuterostome: pelagic or benthic mode of life?

The swimming capacity of acranians now is widely accepted as a plesiomorphic character of the Chordata (since Tunicata are no longer regarded as ancestral forms), and this raises the question whether early deuterostomes were capable of active swimming, too. The idea usually is rejected by authors who favour an enteropneust-like predecessor, and thus a fully benthic mode of life for the ur-deuterostomia (e.g. LACALLI 2005). As mentioned in section 2.2, developmental biologists tend to support this notion because the localization of the ente-ropneust mouth at the chordin-side is protostome-like and thus indicates that enteropneust are links between proto- and deuterostomes (NÜBLER-JUNG & ARENDT 1999; GERHART 2006). However, newer evo/devo-studies realize that the enteropneusts may represent a highly modified deuterostome condition. DENES et al. (2007, p. 285) offer an interpretation which perfectly fits the scenario of the hydroskeleton hypothesis (compare fig. 6 & 7):

> "One possible explanation is that the enteropneust trunk has lost part of its neuroarchitecture due to an evolutionary change in locomotion. While annelids and vertebrates propel themselves through trunk musculature (and associated trunk CNS), the enteropneust body is mainly drawn forward by means of the contraction of the longitudinal muscles in their anterior proboscis and collar (…). Possibly, enteropneusts have partially reduced their locomotor trunk musculature concomitant with motor parts of the CNS (while the peripheral sensory neurons prevailed in "diffuse" arrangement)."

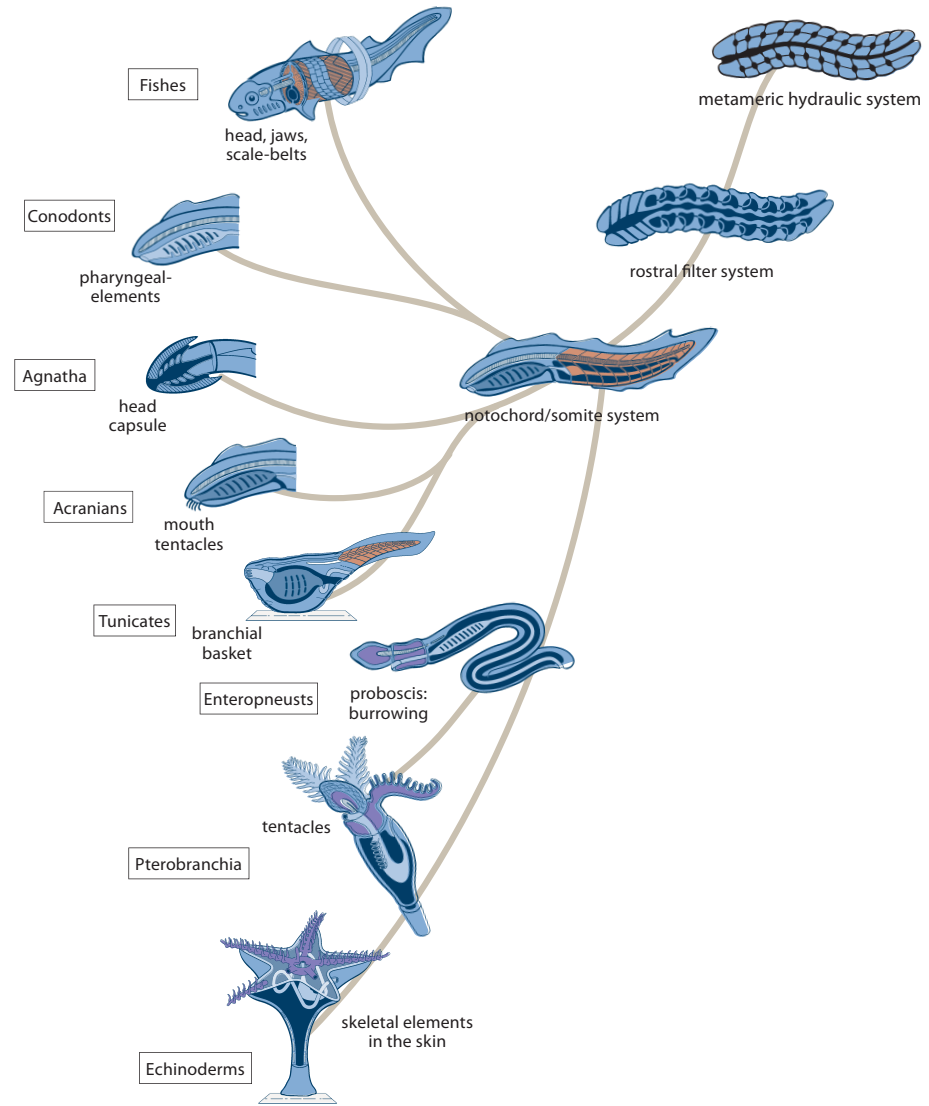

**Fig. 8.** Combined evolutionary scenario (= anagenetic relations) of the deuterostomes. Biomechanical considerations strongly suggest that within the deuterostomes, two major lineages diverged very early: the lineage of the ambulacraria (secondarily benthic forms) and the lineage of the chordata (primarily pelagic; the tunicates secondarily became benthic with the possible exception of the pelagic tunicate subgroup of the Copelata [syn. Larvacea/Appendicularia]). For details, see text.

This justifies the assumption of swimming organisms as protostome-deuterostome intermediates, as more as this seems plausible within the phylogenetic context of the NAP, especially with respect to the systematic position of the Chaetognatha (VARGAS & ABOITIZ 2005). As mentioned above, Chaetognatha appear unrelated to the deuterostomes in molecular phylogenies, and group as an isolated protostome branch ancestral to the Ecdysozoa + Lophotrochozoa-clade. Therefore, VARGAS & ABOITIZ (2005) conclude that ancestral protostomes might well have been pelagic organisms (as chaetognaths typically are), and that this also holds true for the ancestral deuterostomes (even more so as Chaetognatha are "protostomes with deuterostome-like development", MARLÉTAZ et al. 2006). Moreover, BOURLAT et al. (2006: 88) emphasize that the nearest protostome-like relatives of the deuterostomes possess a centralized nervous system rather than diffuse nerve nets, so that the former is more likely to represent the condition of basal deuterostomes (comp. also DENES et al. 2007). Although both the swimming capacity and a centralized nervous system are in agreement with the anagenetic scenario shown in fig. 4, we feel that this argumentation is problematic. For example, VARGAS & ABOITIZ (2005) do not discuss evidence provided by COOK et al. (2004) that Acoelomorpha represent the most basal bila-terian line. Defenders of the "benthic deuterostome ancestor"-hypothesis could interpret the epibenthic lifestyle and the basoepithelial nerve net of acoelomorphs as an ancestral condition, and this could also be claimed for the same features in xenoturbellids (given that they group outside the Deuterostomia, as supposed by PERSEKE et al., 2007). This case exemplifies why instead of parsimony arguments, we prefer detailed anagenetic models to reconstruct the morphological organization and habitat of the urdeuterostomia. According to our argumentation in section 4.2, flatworm-like creeping forms or neotenic larva of such benthic organisms can be ruled out as direct chordate ancestors. For example, it is entirely unclear how a well-regulated muscle/myosept-system could evolve gradually in such forms. GUTMANN (1988, p.265) demonstrated this for the tail of the tunicate larva: These larvae have been claimed to represent chordate ancestors, although no explanation or model has ever been offered for a gradual evolution of myosepta in those organisms. This is also the case in the larva-hypothesis given by SALVINI-PLAWEN (1998, p.138): In his fig. 21, the longitudinal muscles suddenly develop into a serial pattern, but no indication of the causes and constructional consequences of this evolutionary step is provided. The hydroskeleton model does not suffer from such shortcomings, as here the myosepta can be derived from the transverse septa of an annelid-like adult form without unexplained sudden ocurrences of novel structures in the course of the anagenetic development.

### 5.4 Final Conclusions

Anagenetic hypotheses are always present in phylogenetic trees but usually not revealed in full detail, especially in purely cladistic approaches. Following the integration of molecular data and results from the "Konstruktionsmorphologie"-approach as given above for the deuterostome phyla, one can outline a simple, two-step method to evaluate "naked" cladograms (i.e. pure genealogical schemes). The first step is to derive a scenario from the tree considered; the second step is to compare this scenario with anagenetic models that are based on quasi-engineering investigations and reconstructions of the functional designs of the evolutionary stages postulated. In turn, molecular investigations are extremely helpful to determine starting points of anagenetic models, especially for highly modified or miniaturized forms. Above, we have mentioned xenoturbellids, echinoderms and the tunicate subgroup of the copelates as examples how molecular genealogies support the determination of phylogenetic precursor stages. However, secondary reductions and modifications also occur on the genetic level and are difficult to handle by molecular systematics (which actually is a problem in the Tunicata, compare MALLATT & WINCHELL 2007, p.1012). Waiting for robust molecular phylogenies may require considerable patience in some cases, but it is wiser than accepting molecular trees too early, as totally artificial results are always possible. A warning example is the previous molecular-based classification of *Xenoturbella* as a mollusk, more specifically a nuculid clam (literature in PERSEKE et al. 2007). This finding immediately led some authors to suggest that body plan plasticity is nearly incalculable (e.g. LEROI 2000). This view ignores biomechanical constraints and physiological limitations that canalize morphological transformations. Contrarily, in the context of the "Konstruktionsmorphologie"-approach, the lack of any plausible derivation of the *Xenoturbella* body-plan from a clam-like organization forms a strong argument against the validity of the molecular analyses that motivated the inclusion of *Xenoturbella* in the molluscs. Evolution does not stand for a "morphing" of character patterns. Therefore we are skeptical when for example GERHART (2006) presents a character-based body inversion scenario leading from enteropneusts (turned upside down) to chordates. GERHART himself admits on p. 682 that "this is, of course, a speculative >morphing< exercise". In contrast, anagenetic reconstructions that are based on analyses of the biomechanics and physiological functioning of the organisms in question, are open to criticism of their physical plausibility, and – speculative as they may be – can never be rated as mere "morphing exercises" (the latter being a purely statistical approach). We agree with BOCK (2000) who rated process-oriented approaches as "nomological deductive explanations"; such explanations can always be critically evaluated in terms of causations and effects, because they deal with transformations of functional organisms, not with patterns of descriptive characters.

In the case of the deuterostomes, detailed anagenetic reconstructions suggest that all phylogenetic trees are highly questionable which, for example, shift echinoderms to an ancestral position, or which combine echinoderms or hemichordates with tunicates (known as the pharyngopneust-hypothesis). Similarly, assumptions of a xenoturbellid-like or an oligomeric (trimeric) ur-deuterostomian have to be rejected: The mechanical conditions for the development of a notochord and an efficient locomotory apparatus as known from recent acranians and craniotes, are not given in any of these body structures. On the other hand, trees are plausible that place echinoderms and hemichordates together as Ambulacraria, as the sister group of the Chordata. Highly convincing are those phylograms which suggest a polysegmented ancestor for the deuterostomes (as for example fig. 1), because only in a polysegmented system the biomechanical conditions for the development of a notochord are present.

What does this imply regarding the Deuterostomia-phylogenies (as shown in fig. 2b) of the NAP? Even though there still is some ambiguity in the cladogenetic model, the overall anagenetic implications appear quite plausible from a constructional morphologist´s point of view, especially when results from molecular systematics and developmental genetics are combined, as shown in fig. 1. In other words, the Deuterostomia-phylogenies elaborated by newer molecular approaches can be rated as being in the realm of possibility. If so, GROBBEN'S (1908) structure-based phylogeny, deriving deuterostomes from coelomates and deviding them into Ambulacraria and Chordata implies one of the most favourable anagenetic models – however, a lot of details of this scenario still await plausible reconstructions.

## 6. Acknowledgements

We are indebted to our colleagues and students at the Senckenberg Institute for discussions of the topic. Special thanks to Dr. MANFRED GRASSHOFF, who read the manuscript several times and helped considerably to develop the argumentation presented, and to Dr. WINFRIED S. PETERS for improving the manuscript linguisticly.